\documentclass[journal]{IEEEtran}
\usepackage{graphicx}
\usepackage{amsmath}
\usepackage{subfigure}
\usepackage{multirow}
\usepackage{array}
\ifCLASSINFOpdf

\else

\fi

\begin{document}
%
\title{Modeling Routing Overhead Generated by \\ Wireless Reactive Routing Protocols}

\author{\IEEEauthorblockN{Nadeem Javaid, Ayesha Bibi, Akmal Javaid, Shahzad A. Malik\\\vspace{0.4cm}}
        Department of Electrical Engineering, COMSATS Institute of\\
        Information Technology, 44000, Islamabad, Pakistan. \\
        nadeemjavaid@comsats.edu.pk\\

     }
\vspace{-2cm}

%

\maketitle

\begin{abstract}
In this paper, we have modeled the routing overhead generated by three reactive routing protocols; Ad-hoc On-demand Distance Vector (AODV),
Dynamic Source Routing (DSR) and DYnamic MANET On-deman (DYMO). Routing performed by reactive protocols consists of two phases; route discovery
and route maintenance. Total cost paid by a protocol for efficient routing is sum of the cost paid in the form of energy
consumed and time spent. These protocols majorly focus on the optimization performed by expanding ring
search algorithm to control the flooding generated by the mechanism of blind flooding. So, we have modeled
the energy consumed and time spent per packet both for route discovery and route maintenance. The proposed framework is evaluated in NS-2 to compare performance of the chosen routing protocols.

\end{abstract}

\begin{IEEEkeywords}
Routing protocols, expanding ring search, routing overhead, AODV, DSR, DYMO, route discovery, route maintenance
\end{IEEEkeywords}

\IEEEpeerreviewmaketitle

\vspace{-0.2cm}
\section{Introduction}

\IEEEPARstart{R}{ECENT} era has seen the plenty of mobile networks due to the proliferation of wireless devices all around. Specially, recent studies have mainly focused on Wireless Multi-hop Networks (WMhNs). In a WMhN, every node can act both as a sender/receiver and as a router for the nodes which are not in the same transmission range. These kind of networks allow a vast range of applications ranging from a conference or a business meeting to the environments like a battlefield or a natural disaster scenario. Performance of all WMhNs is majorly depending on the routing protocols operating them.

Routing protocols are of two types with respect to their routing nature; reactive and proactive. The main difference between the two categories is based on the calculation of routes. Proactive protocols periodically perform routing table calculation independently from the data request arrival; like DSDV [1], FSR [2], [3], OLSR [4], [5], etc. Reactive protocols calculate routes on reaction of data request arrives; rightly called 'on-demand' routing protocols; e.g., AODV [6], DSR [7], [8], TORA [9], DYMO [10], [11], etc.

Reactive protocols are best suited for the networks with higher rates of mobility and proactive ones are designed for dense and static networks. To keep up-to-date all nodes about all links, reactive protocols have to deal with frequent link breaks due to high mobilities. For this, reactive protocols exchange lot of control (routing) packets increasing routing overhead. In this paper, we, therefore, have modeled the routing overhead generated by the above mentioned three reactive routing protocols. They perform route discovery $(RD)$ and route maintenance $(RM)$ operations for routing. In the earlier process, the protocols find the route for desired destination, while the former process starts after establishment of routes; as, monitoring the active routes and appropriate actions to be taken after link breakage detection. Expanding Ring Search (ERS) algorithm and Binary Exponential Back-off (BEB) algorithms are used by all of the three protocols with different parameters, as, shown in Fig.1. Gratuitous Route REPlies \textit{(Grat. RREPs)} are used by DSR and AODV during $RD$ process unlike DYMO to reduce routing overhead. During route repair, DSR adopts packet salvaging $(PS)$ due to promiscuous listening mode, while AODV implements local link repair $(LLR)$ after link break detection.

Rest of the paper is arranged as follows: sectionII discusses related work and motivation. SectionIII and IV describe optimization of flooding
using ERS algorithm. The detailed framework for routing operations in reactive protocols is explained in sectionV.  Moreover, for testing the proposed framework in NS-2, sectionVI gives analytical simulation results.

\vspace{-0.2cm}
\section{Related Work and Motivation}
In [12], authors make a survey about control overhead of ad-hoc routing protocols. They characterize reactive and proactive protocols as "hello protocols" and "flooding protocols" and prove from simulations that more control packets are needed for hello protocols in mobile scenarios as compared to flooding protocols. Therefore, hello protocols are more suitable for fixed scenarios and flooding protocols are well suited for mobile situations. However, Jacquet, P. \textit{et. al} [12] only discuss energy cost for routing protocols. While, we model both energy and time costs for reactive protocols.

Authors in [13] give an expression $C_{total}^{(rp)}=C_{E}^{(rp)}\times C_{T}^{(rp)}$, for measurement of total cost; $C_{total}^{(rp)}$ of a routing protocol which is the product of energy cost; $C_{E}^{(rp)}$ and time cost; $C_{T}^{(rp)}$. While $C_{E}^{(rp)}$ means energy consumed while $C_{T}^{(rp)}$ is the time spent by a protocol for routing operations; $RD$ and $RM$. Authors compare simple flooding with ERS basic mechanism. Moreover, the overhead cost is evaluated as the total cost. Whereas, we further classify both time cost and energy cost for $RD$ and $RM$. Therefore, $C_E^{(rp)}=C_{(E-RD)}^{(rp)}+C_{(E-RM)}^{(rp)}$ and $C_T^{(rp)}=C_{(T-RD)}^{(rp)}+C_{(T-RM)}^{(rp)}$ are the respective energy and time cost expressions for a routing protocol in our framework. Moreover, we relate these costs not only with basic cost of ERS for AODV, DSR and DYMO, but also we present improved equations for ERS.

\begin{figure}[!h]
\begin{center}
\includegraphics[
height=9 cm,
width=8 cm
]{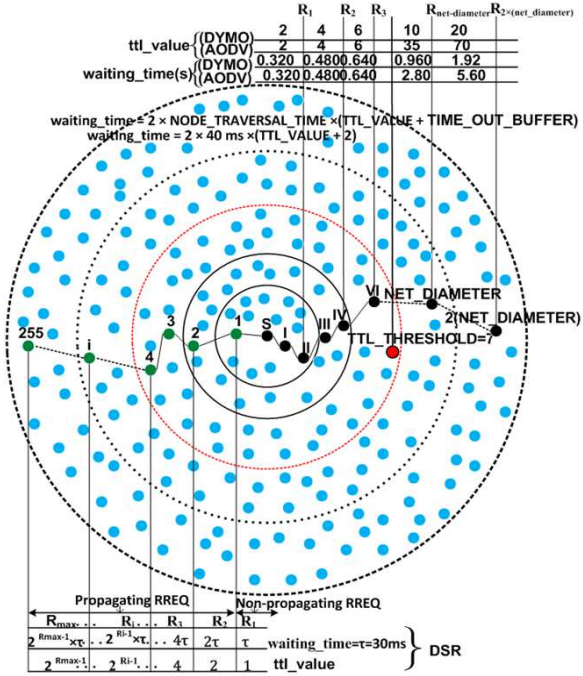}
\end{center}
\caption{ERS Algorithm used by AODV, DSR, DYMO}
\end{figure}

Lin, T. \textit{et. al} [14] present an analytical model for comparison of routing protocols proposing a framework with overhead as a metric. They compare the use of relay nodes of proactive protocols with flooding process of reactive protocols. Whereas, we consider time cost along with routing overhead and also make intra-comparison model for reactive protocols by presenting separate models of respective flooding mechanisms along with the optimization strategies.

Saleem \textit{et. al} [15] improve their work in [16], by taking inspiration from Broch, J. \textit{et. al} [17] and present flooding cost of routing protocols. They present a performance evaluation framework that can be used to model two key performance metrics of an ad-hoc routing algorithm; \textit{routing overhead} and \textit{route optimality}. They compare two metrics; \textit{total energy consumption} and \textit{route discovery latency} to evaluate the performance of four prominent ad-hoc routing algorithms; DSDV, DSR, AODV-LL and Gossiping. However, they do not specify any cost for $RM$ process for reactive protocols. They also do not model any of the supplementary strategies of routing protocols and expression for time consumption.  While, we consider broadcast probability; $P_s$ equal to stochastic forwarding; $P_r$, in the same way as specified in [16]. We not only model the routing packet cost for AODV, DSR and DYMO, but also time expenditure for these protocols during their routing operations. Then we consider supplementary strategies (\textit{grat. RREPs}, $LLR$ and $PS$) of respective protocols to accurately evaluate their routing costs.

\vspace{-0.2cm}
\section{Flooding in Reactive Protocols}
A flooding algorithm is used for exchanging the topological information with every part of a network. In flooding each node can act both as a source and as a router. Each node broadcasts route information to all of its neighbors until destination is reached. This repeated broadcast results in the reception of a particular message by all nodes in the network. A straightforward approach for broadcasting as flooding technique is \textit{blind flooding}; each node is required to rebroadcast the packet whenever it receives the packet for the first time. \textit{Blind flooding} can cause the broadcast storm problem by generating the redundant transmissions.

Each routing protocol has to pay some cost for the routing overhead because of flooding. In \cite{16}, authors present following equation.

\small
\begin{eqnarray}
 C_p=
  \begin{cases}
   P_Sd_{avg} & if \,\, h=1 \\
   P_Sd_{avg}+d_{avg}\displaystyle\sum_{i=1}^{h-1}(P_S)^{i+1}\prod_{j=1}^{i}d_f[j]       & otherwise
  \end{cases}
\end{eqnarray}
\normalsize

The number of directly connected or adjacent neighbor nodes within a network for a node is known as \textit{degree} of that node. An isolated node is a node which has zero \textit{degree}. In eq.1, $h$ is the number of hops, $d_{avg}$ is average degree of a node, $d_f[j]$ is the expected forward degree of a node at the $j^{th}$ hop and is the average number of neighbors of that node which forwards a received RREQ with probability of broadcasting; $P_S$ [18].

\vspace{-0.2cm}
\section{Optimization of Flooding using ERS Algorithm}
Eq.1 gives an approximate cost paid by a protocol per packet for $RD$ using \textit{blind flooding}. There are many optimizations to control the routing overhead. ERS [19] is one of the optimization techniques and is adopted by AODV, DSR and DYMO. In ERS, the flooding is controlled by the Time To Live (TTL) values to limit the broadcast.

As, ERS uses \textit{blind flooding} for broadcasting, so? its routing cost can be calculated from eq.1. In the case of ERS, $h$ is replaced by the TTL value in a ring. $C_{E-R_i}$ is the cost of any ring; $R_i$ that generates RREP(s) and the ring $R_i$ is called $R_{rrep}$ and it can disseminate up to the maximum limit $R_{max\_limit}$ resulting in either successful or unsuccessful $RD$; i. e; $R_i\backslash R_i\in R_{rrep}\vee R_i\in R_{max\_limit}$. So, $C_{E-R_i}$ can be calculated as:


\tiny
\begin{eqnarray}
 C_{E-R_i}=
  \begin{cases}
   P_r d_{avg}  & if\,  TTL(R_i)=1 \\ \nonumber
   P_r d_{avg}+d_{avg}\displaystyle\sum_{TTL=1}^{TTL(R_i)-1}(P_r)^{TTL+1}\prod_{j=1}^{TTL}d_f[j] & otherwise \,\,\,\,\,(1a)\\
   \{R_i\backslash R_i\in R_i\rightarrow R_{rrep}\vee R_i\in R_i\rightarrow R_{max\_limit}\}
  \end{cases}
\end{eqnarray}
\normalsize

\tiny
\begin{eqnarray}
 C_{E-RD}=
  \begin{cases}
   \displaystyle\displaystyle\sum_{R_i=1}^{R_{max\_limit}}(C_{E-R_i})_{R_i} & if\, no\, RREP\, received \\
   C_{E-R_{rrep}} & if\, TTL(R_{rrep})=1 \\
   {\displaystyle\displaystyle\sum_{R_i=1}^{R_{rrep}}(C_{E-R_i}})_{R_i} & otherwise\\
   \{R_{rrep}=1,2,3,....,max\_limit\}
  \end{cases}
\end{eqnarray}
\normalsize

$RD$ using ERS requires broadcast inside the rings by incrementing TTL values relative to the previous TTL value. In ERS, gradual growth of broadcasting ring takes place to reduce the chances of flooding in the entire network that results in formation of different numbers of rings for different broadcasting levels, as shown in Fig.1. The collective routing cost of these expanding rings during $RD$ process $C_{E-RD}$ can be computed using eq.2.

\vspace{-0.2cm}
\section{Modeling the cost paid by reactive protocols}
A routing protocol, $p$ has to pay some cost in the form of consumed energy per packet, $C_E^{(p)}$ and in the form of time spent per packet, $C_T^{(p)}$ to encounter the topological changes during the varying number of nodes and traffic rates. In [13], the authors have expressed this cost by the following equation:

\small
\begin{eqnarray}
C_{total}^{(p)}=C_{E}^{(p)} \times C_{T}^{(p)}
\end{eqnarray}
\normalsize

\subsection{Cost of Energy Consumption}
Each reactive protocol, $rp$ performs $RD$ and $RM$, so, unlike, eq.3, we define the cost to be paid for energy consumption during $RD$ and $RM$ processes; $C_E^{(rp)}$:

\small
\begin{eqnarray}
C_{E}^{(rp)}=C_{E-RD}^{(rp)} + C_{E-RM}^{(rp)}
\end{eqnarray}
\normalsize

$C_E^{(rp)}$ is different for each reactive protocol due to different routing strategies. The multiple routes in Route Cache \textit{(RC)} reduce the routing overhead with the help of \textit{grat. RREPs} and \textit{PS} in DSR. In AODV, route length is shortened by \textit{grat. RREPs} to reduce the cost of \textit{RD} process while successful \textit{LLR} probability; $P_S^{LLR}$ avoids the re-initiation of \textit{RD} process.

\textbf{Energy Consumed for \textit{RD}}

AODV, DSR and DYMO use ERS for $RD$ by broadcasting the RREQ messages from the source node. A source node may receive RREPs from the nodes that contain alternate (short) route for the desired destination. These replies are only used in AODV and DSR and are known as \textit{grat. RREPs}. The destination RREPs are generated by the destination itself (destination RREPs are generated in all the three reactive protocols).

Eq.5 gives the cost to be paid for RREQ packets as well as the cost for RREPs produced during $RD$. $n_{rrep}$ notation is used for node(s) generating the RREP from $R_{rrep}$.

\tiny
\begin{eqnarray}
 C_{E-RD}^{(rp)}=
  \begin{cases}
   \displaystyle\sum_{R_i=1}^{R_{max\_limit}}(C_{E-R_i})_{R_i} & if\, no\, RREP\, received \\
   C_{E-R_{rrep}}+ \displaystyle\sum_{n=1}^{n_{rrep}}(RREP)_n & if\, TTL(R_{rrep})=1 \\
   \displaystyle\sum_{R_i=1}^{R_{rrep}}(C_{E-R_i})_{R_i}+ \displaystyle\sum_{n=1}^{n_{rrep}}(RREP)_n & otherwise\\
   \{R_{rrep}=1,2,3,....,max\_limit\}
  \end{cases}
\end{eqnarray}
\normalsize

The generation of RREP(s) in AODV and DSR is also due to the valid routes in Routing Table $(RT)$ or $(RC)$, so, $R_i$ for DSR and AODV is less than DYMO, as, \textit{grat. RREPs} are absent in DYMO. $rrep$ in $R_{rrep}$ can be written as: $rrep=1,2,3,....,max\_limit$.
Whereas, $d_{rrep}$ represents RREP generated by destination.


\textbf{Energy Consumed for \textit{RM}}

In $RM$ process, different protocols pay different costs for link monitoring and also there are different costs for different supplementary maintenance strategies in case of link breakage; as, $C_{E-LLR}$ for AODV, $C_{E-PS}$ for DSR, and DYMO does not use any mechanism. Following equations give $RM$ cost for three protocols:

\small
\begin{eqnarray}
C_{E-RM}^{(AODV)}=C_{E-l-mon}+C_{E-LLR}
+\displaystyle\sum_{z=0}^{n}(RERR)_z &
\end{eqnarray}
\normalsize

\small
\begin{eqnarray}
C_{E-RM}^{(DSR)}=C_{E-PS}+\displaystyle\sum_{z=0}^{n}(RERR)_z &
\end{eqnarray}
\normalsize

\small
\begin{eqnarray}
C_{E-RM}^{(DSR)}=\displaystyle\sum_{k=n_{BLB}}^{n_{PS}}(RREQ)_k+\displaystyle\sum_{z=0}^{n}(RERR)_z &
\end{eqnarray}
\normalsize

\small
\begin{eqnarray}
C_{E-RM}^{(DYMO)}=C_{E-l-mon}+\displaystyle\sum_{z=0}^{n}(RERR)_z &
\end{eqnarray}
\normalsize

Where, $C_{E-l-mon}$ is the link monitoring cost paid by AODV and DYMO in the form of HELLO messages and it is given in eq.10.

\small
\begin{eqnarray}
C_{E-l-mon}=\frac{\tau_{link-in-use}}{\tau_{HELLO\_INTERVAL}}\times N_{hops-in-route}
\end{eqnarray}
\normalsize

Where, $n_{BLB}$ is the node before link break and $n_{PS}$ may be any node from source to $n_{BLB}$. In wireless environment, there are frequent link breakages that lead to link failures. As a result, the routes become ineffective. The link breakage is detected by different protocols by different strategies. DYMO and AODV generate HELLO messages to check the connectivity of active routes, while DSR gets the link level feedback from link layer. This cost depends upon the link time, i.e. a link in use (active link); $\tau_{link-in-use}$ and length of the path in hops and the value of $HELLO\_INTERVAL$ constant. Broadcast needs to send $z$ number of RERRs depending upon different situations for different protocols, as discussed below.

In DYMO, link breakage causes the broadcasting of RERR messages. When the probability of successful for $LLR$; becomes zero then it leads to the dissemination of RERRs in AODV.

On the other hand, DSR piggybacks RERR messages along with next RREQs in the case of route re-discovery process, while these RERR messages are broadcasted in the case of success of $PS$.


There are three different scenarios for reactive protocols describing the route repair mechanism after detection of route failure because of link breakage. The most simple mechanism describes that $RD$ re-initiation process takes place under the limited retries constraint for route re-discovery process:

$RREQ\_RETRIES = 3$ in DYMO,

$RREQ\_TRIES = 2$ in AODV, and

$MaxMainRexmt$ in DSR=2 retransmissions.

In AODV after unsuccessful $RD$ and normally in DYMO, RERR messages are broadcasted by the node which detects any link break and route rediscovery process is started through source node. Fig.2.a. clearly demonstrates the importance of $LLR$ in AODV. If it becomes successful in a dense network then it saves the energy consumed during route re-discovery. On the other hand, if $LLR$ becomes unsuccessful then energy cost is increased by re-initiating $RD$ process after performing $LLR$ strategy, as depicted in Fig.2.b.

DSR's $PS$ technique can reduce both the energy and time cost to be paid by a reactive protocol by diminishing the route re-discovery. In the case of successful $PS$, RERR messages are broadcasted to neighbors for the deletion of useless routes. Whereas, the absence of alternate route(s) in $RC$ leads to the failure of $PS$. In this situation, RERR messages are to be sent by piggybacking them in the next RREQ messages during route re-discovery process .

Cost of $LLR$ in AODV is given by the following equation.

\tiny
\begin{eqnarray}
 C_{E-LLR}^{(AODV)}=P_rd_{avg}+d_{avg}\displaystyle
 \sum_{TTL=1}^{TTL(R_{LLR})-1}(P_r)^{TTL+1}\prod_{j=1}^{TTL}d_f[j]
 \end{eqnarray}
 \normalsize

Here, $R_{LLR}$ represents the ring that limits $LLR$ activity. TTL value for $R_{LLR}$ is calculated with $LOCAL\_ADD\_TTL(=2)$ and $MIN\_REPAIR\_TTL$ (it is last known hop-count to the destination). The per packet cost of $LLR$; $C_{E-LLR}^{(AODV)}$ depends upon the TTL value of $R_{LLR}$. In large networks, successful $LLR$ process is more useful, because the chances of route re-discovery can be reduced which utilizes more bandwidth space. $TTL(R_{LLR})$ is obtained from the equation given below:

\tiny
\begin{eqnarray}
max(MIN\_REPAIR\_TTL, 0.5 \times \#hops) + LOCAL\_ADD\_TTL
\end{eqnarray}
\normalsize

Where $\#hops$ is the number of hops to the sender of the currently undeliverable data packet. Thus, local repair attempts will often be imperceptible to the originating node, and will always have
$TTL >=MIN\_REPAIR\_TTL + LOCAL\_ADD\_TTL$.

\subsection{Cost of Time Consumption}
The cost of end-to-end path calculation time $C_T^{(rp)}$ in reactive protocols depends upon $C_{T-RD}^{(rp)}$ and $C_{T-RM}^{(rp)}$.

\small
\begin{eqnarray}
C_T^{(rp)}=C_{T-RD}^{(rp)}+C_{T-RM}^{(rp)}
\end{eqnarray}
\normalsize

\textbf{Time Consumed for \textit{RD} by DSR}

$\tau$ is constant time initially used for non-propagating RREQ $(NonpropRequestTimeout)$ and its value is $30ms$. $R_{max\_limit}$ is the maximum ring size and it depends on the buffer time as well as the maximum allowed broadcasting during propagating RREQ $(DiscoveryHopLimit=255)$. The BEB is associated with each propagating ring. The expression for Time Consumed for $RD$ by DSR is given below.

\small
\begin{eqnarray}
 C_{T-RD}^{(DSR)}=
  \begin{cases}
  \tau & if \,R_{rrep}=1 \\
   \displaystyle\sum_{{R_i}=1}^{R_{max\_limit}}2^{{R_i}-1}\times \tau  & if \,no\,RREP\,received \\
    \displaystyle\sum_{{R_i}=1}^{R_{rrep}}2^{{R_i}-1}\times \tau  & otherwise
  \end{cases}
\end{eqnarray}
\normalsize

\textbf{Time Consumed for \textit{RD} by AODV and DYMO}

Both in AODV and DYMO, firstly, $TTL\_VALUE$ in IP header is set to $TTL\_START$ (=1 in the case of link layer feedback otherwise =2) then it is increased by $TTL\_INCREMENT (=2)$  up to $TTL\_THRESHOLD (=7)$ \cite{6}. When $TTL\_THRESHOLD$ is reached, $TTL\_VALUE$ is set to $NET\_DIAMETER$ (for AODV = 35 \cite{6} and for DYMO = 10 \cite{7}). For dissemination in the entire network, $TTL\_START$ and $TTL\_INCREMENT$ both are set to $NET\_DIAMETER$. Moreover, maximum RREQ tries are $3$ for DYMO \cite{7}, and maximum retries are $2$ for AODV. The $RREQ\_TIME$ is set to $2 \times NET\_TRAVERSAL\_TIME$ (whereas, $NET\_TRAVERSAL\_TIME=2\times NODE\_TRAVERSAL\_TIME\times NET\_DIAMETER)$.

\tiny
\begin{eqnarray}
 C_{T-RD}^{(AODV,\,DYMO)}=
  \begin{cases}
\displaystyle\sum_{R_i=1}^{R_{max\_limit}}\tau_1(TTL(R_i)+\tau_2)  & if \,no\,RREP\,received \\\\ \nonumber
\displaystyle\sum_{R_i=1}^{R_{rrep}}\tau_1(TTL(R_i)+\tau_2)  & otherwise \,\,\,\,\,\,\,\,\,\,\,\,\,\,\,\,\,\,\,\,\,\,\,(14a) \\
  \end{cases}
\end{eqnarray}
\normalsize

Where, $\tau 1=2\times NODE\_TRAVERSAL\_TIME$ and $\tau 2=TIME\_OUT\_BUFFER$. There are two possibilities for AODV and DYMO; first is the case when $RD$ process becomes successful in threshold rings $R_{thereshold}$, while in second case $RD$ process needs to disseminate the request in the whole network; $R_{netdiameter}$. For these two rings, we define $TTL(R_{thrshold})$ and $TTL(R_{netdiameter})$. Earlier represents the TTL value in a ring which generates RREP(s) inside $R_{thereshold}$ with $THRESHOLD$ and later shows TTL vale for the entire network; $TTL(R_{netdiameter})$ with $NET\_DIMETER$.

\textbf{Time Consumed for \textit{RM} in AODV}

AODV starts $LLR$ process after noticing a link failure. $C_{T-LLR}$ gives the time cost of $LLR$ that depends upon the TTL value of the ring; $LLR(R_{LLR})$. In the case of $LLR$ failure, AODV disseminates RERR messages. $\tau_{recv-RERR}$ represents the time which is spent to reach RERR message from the node detecting the link failure to the originator node. $C^{(AODV)}_{T-re-RD}$ cost is to be paid to start route re-discovery based on the value $RREQ\_RETRIES(=2)$.

\small
\begin{eqnarray}
 C_{T-RM}^{(AODV)}=
  \begin{cases}
A & if \,LLR\,is\,successful\\
B & if \,LLR\,fails,\,RREQ\_RETRIES\,expires \\
C & otherwise \\
  \end{cases}
\end{eqnarray}
\normalsize

where,

$A=\displaystyle\sum_{R_i=1}^{R_{LLR}}\tau_1(TTL(R_i)+\tau_2)$\\

$B=\displaystyle\sum_{R_i=1}^{R_{LLR}}\tau_1(TTL(R_i)+\tau_2)+\tau_{recv-RERR}$\\

$C=\displaystyle\sum_{R_i=1}^{R_{LLR}}\tau_1(TTL(R_i)+\tau_2)+\tau_{recv-RERR}+C_{T-re-RD}$\\

$
C^{(AODV)}_{T-re-RD}=C^{(AODV)}_{T-RD}
$

\textbf{Time Consumed for \textit{RM} in DSR}

After detecting a link failure, time $\tau_{PS}$ is utilized to check alternative routes in $RC$ of intermediate nodes (from a node which detects link failure to a node having alternate route for this broken link; $n_{PS}$. This $\tau_{PS}$ value is higher; if node containing alternative route; $n_{PS}$ is nearest to the node which detects link breakage. In the case of failure of $PS$ or in the case of presence of alternative route in $RC$ of the originator node, $\tau_{PS}$ attains a maximum value and is consumed by all intermediate remaining nodes in a route (from a node that detects link break up to the originator of this broken route).

\tiny
\begin{eqnarray}
C_{T-RM}^{(DSR)}=
\begin{cases}
\displaystyle\sum_{k=n_{BLB}}^{n_{PS}}\tau_k(PS) & if \,PS\,is\,successful\\
\displaystyle\sum_{k=n_{BLB}}^{n_{orig}}\tau_k(PS)+C^{(DSR)}_{T-re-RD} & otherwise
 \end{cases}
\end{eqnarray}
\normalsize

where, $C^{(DSR)}_{T-re-RD}=C^{(DSR)}_{T-RD}$, $n_{BLB}$ is the node just before link breakage, and $n_{orig}$ is the node which originates route discovery process.

\textbf{Time Consumed for \textit{RM} in DYMO}

A RERR message is broadcasted by the node that detects link break. After a time $\tau_{recv-RERR}$, which is consumed for receiving RERR message by the source node, source node initiates $RD$; $C^{(DYMO)}_{T-re-RD}$ is based on $RREQ\_RETRIES(=3)$ constraint.

\tiny
\begin{eqnarray}
 C_{T-RM}^{(DYMO)}=
  \begin{cases}
\tau_{recv-RERR}& if \,RREQ\_TRIES\,expires \\
\tau_{recv-RERR}+C^{(DYMO)}_{T-re-RD}& otherwise \\
  \end{cases}
\end{eqnarray}
\normalsize

\textbf{Time-Trade-Offs between \textit{PS} and \textit{LLR}}

In $PS$ of DSR, checking the $RC$ of intermediate nodes for alternate route(s) consumes more time. In the case of successful $PS$, time can be reduced as compared to time consumption for route re-discovery from source for the end-to-end path calculation. On the other hand, $PS$ checking time adds-up with route re-discovery time in case of failure.

Same is the case with $LLR$ in AODV. Success of the process lessens end-to-end path time because route re-rediscovery process is not initialized. While, $LLR$ increase the path length in the case of unsuccessful repair, because the repair time is also added with re-discovery time. As, $LLR$ is performed by broadcasting a small TTL value, so, $R_{LLR}$ consumes some time during repair. The time cost of $LLR$ in AODV $C_{T-LLR}^{(AODV)}$ can be calculated as:

\small
\begin{eqnarray}
C_{T-LLR}^{(AODV)}= \displaystyle\sum_{R_i=1}^{LLR}\tau_1(TTL(R_i)+\tau_2)
\end{eqnarray}
\normalsize

\vspace{-0.2cm}
\section{Analytical Simulation Results Corresponding to Designed Framework}
We evaluate performance of our modeled framework in NS-2. For simulation setup, we have chosen Continuous Bit Rate (CBR) traffic sources with a packet size of 512 bytes. The mobility model used is Random Waypoint. We propose three different scenarios:

\begin{figure}[h]
  \centering
 \subfigure[Control Overhead for 5 hops]{\includegraphics[height=2 cm,width=4.3 cm]{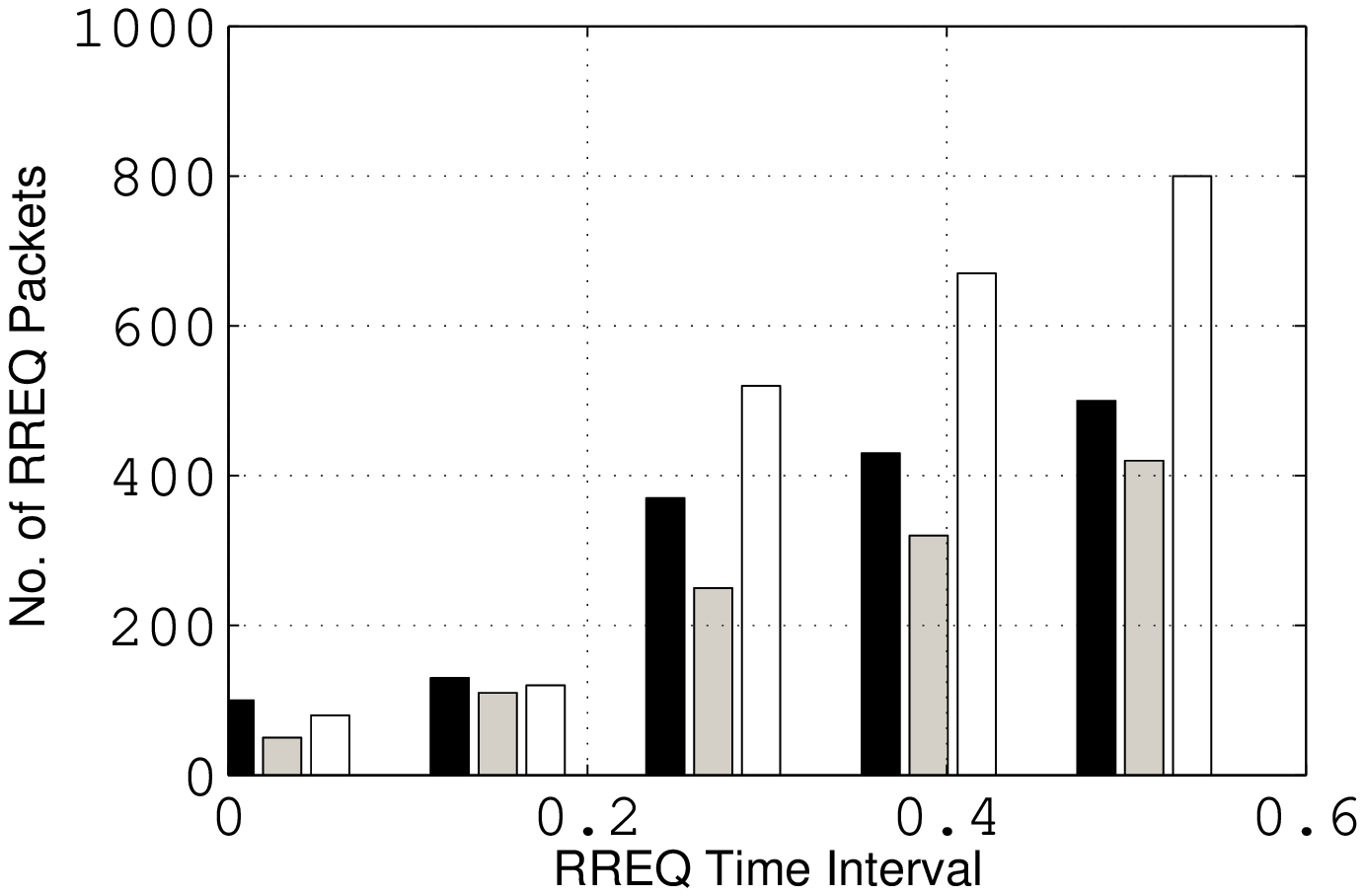}}
 \subfigure[Control Overhead for 10 hops]{\includegraphics[height=2  cm,width=4.3 cm]{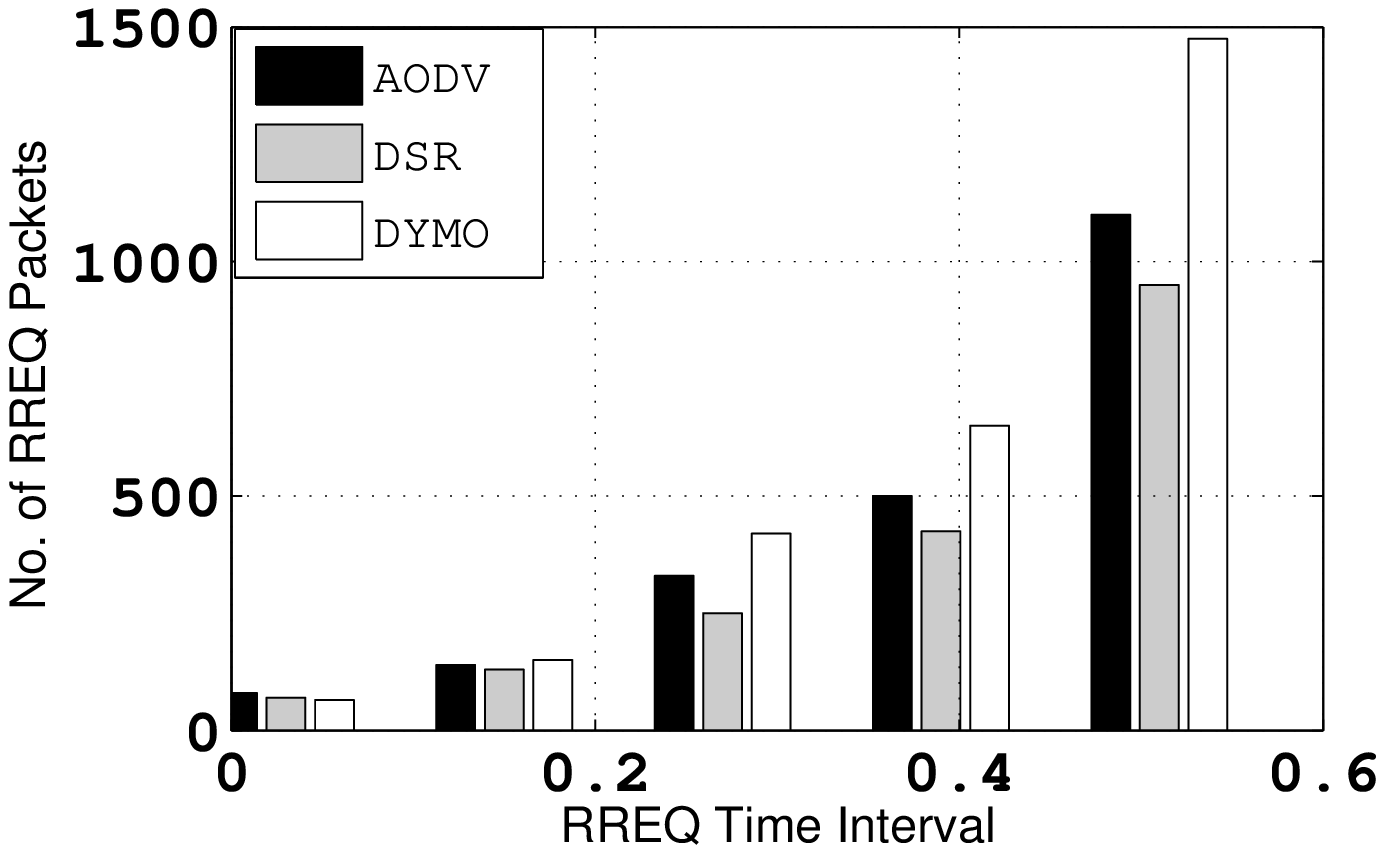}}
  \caption{Control Overhead generated by reactive protocols}
\end{figure}

\textbf{Scenario.1}
Area for simulation is $800m \times 600m$. 50 nodes are moving with speed of 5m/s for simulation time of 300s with 2s pause time. Scenario.1, as depicted in Fig.2, is performed by counting the distance measured by routing protocols in hops to construct appropriate path. As, wireless nodes also behave as routers for the nodes which are not in the same transmission range, it is more appropriate to count the distance in hops.

\begin{figure}[h]
  \centering
 \subfigure[E2ED for higher mobilities]{\includegraphics[height=2 cm,width=4.3 cm]{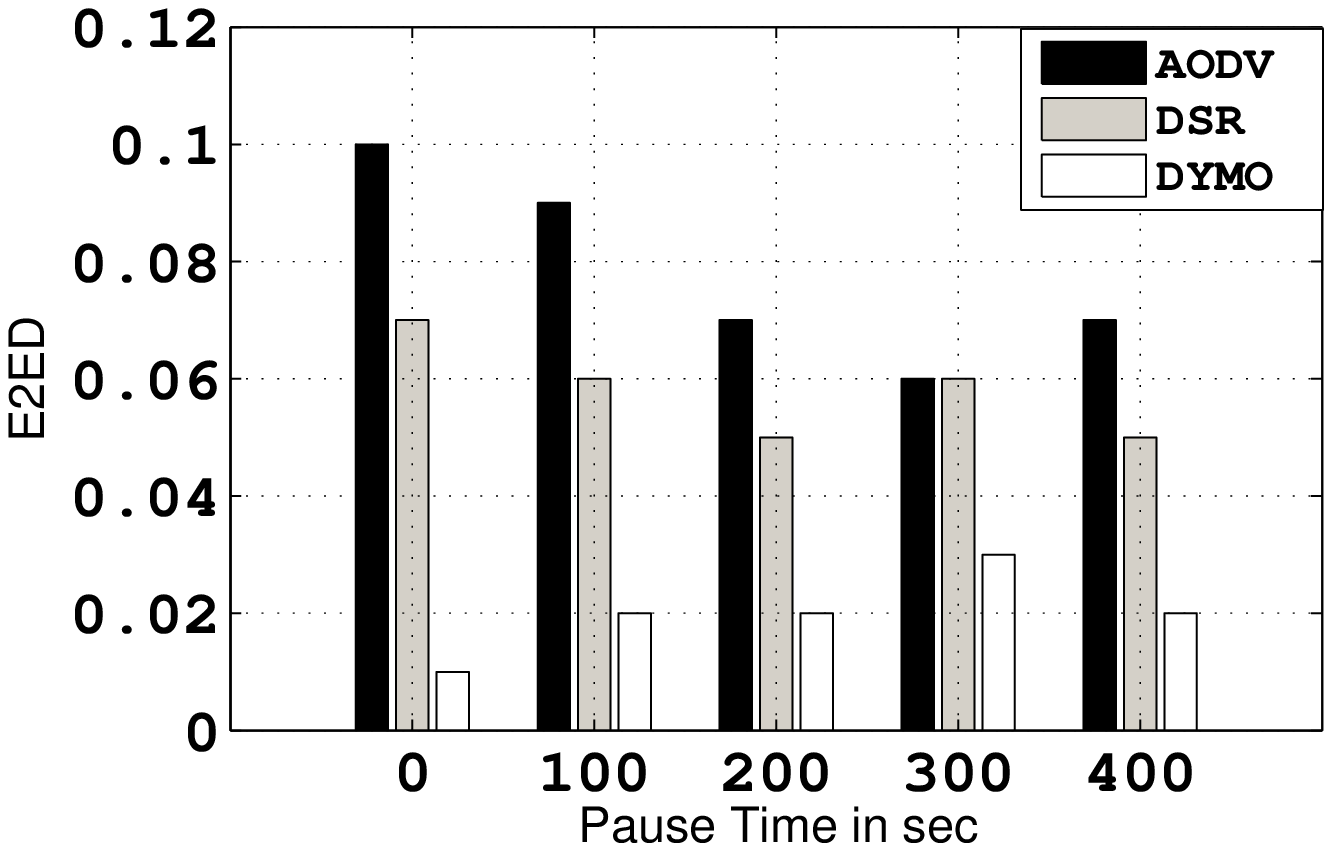}}
 \subfigure[E2ED for lower mobilities]{\includegraphics[height=2  cm,width=4.3 cm]{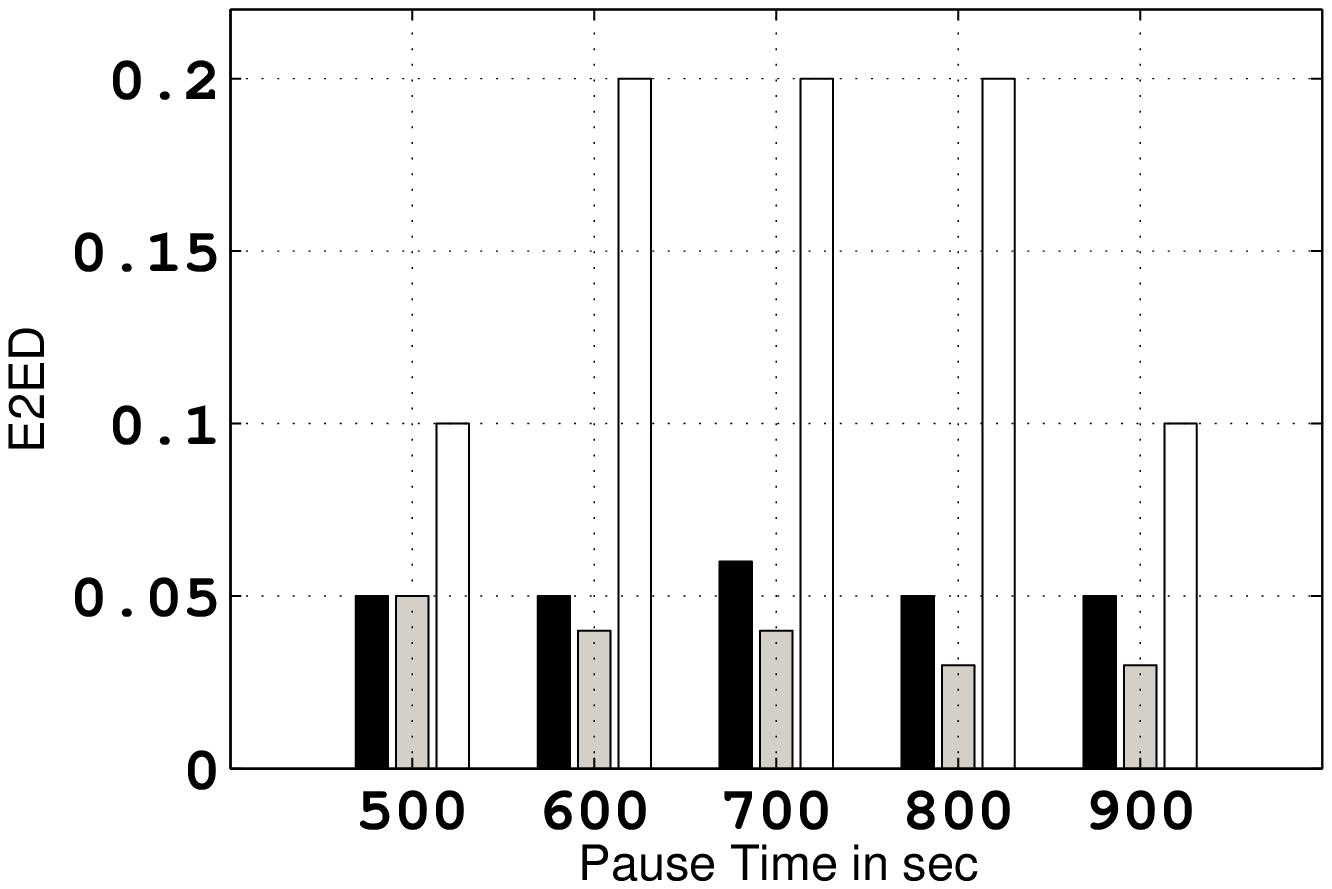}}
 \subfigure[NRL for higher mobilities]{\includegraphics[height=2 cm,width=4.3 cm]{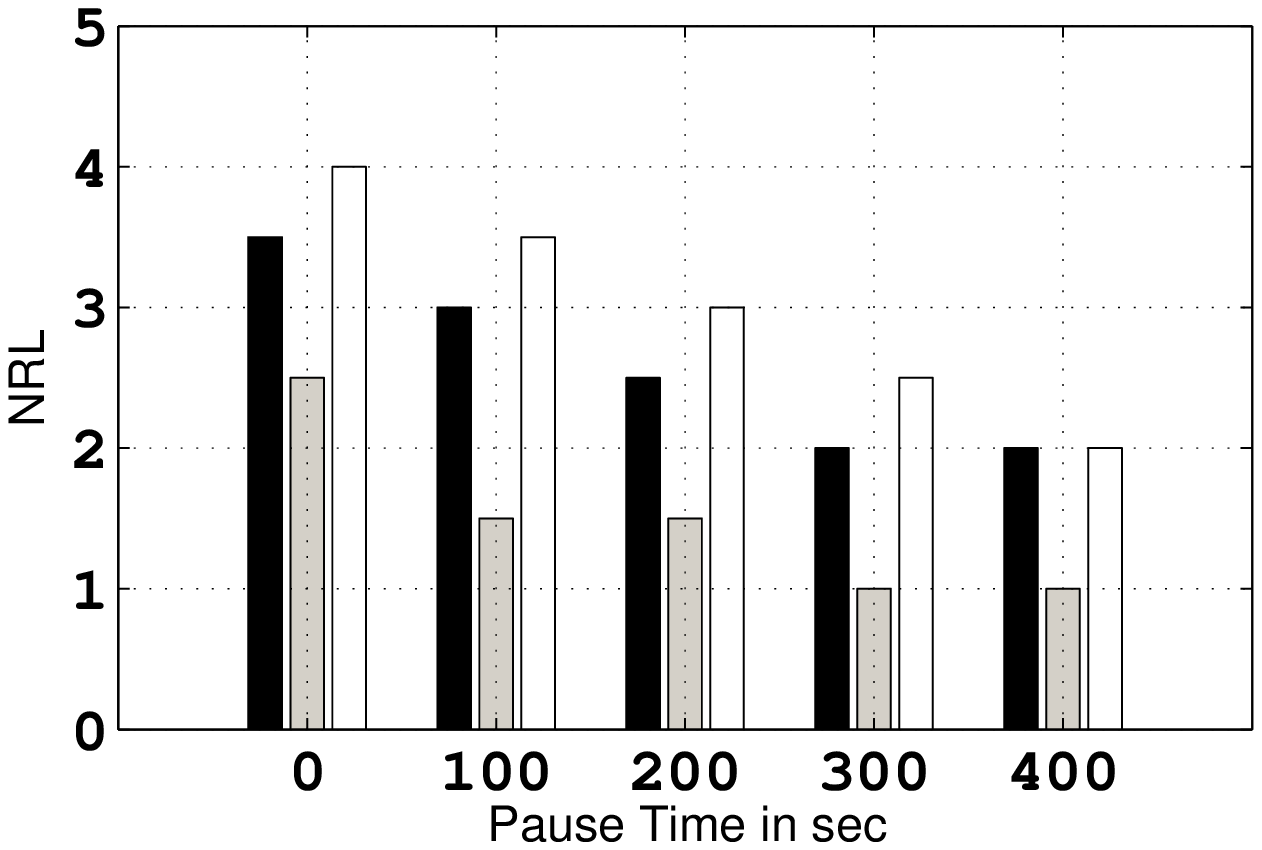}}
 \subfigure[NRL for lower mobilities]{\includegraphics[height=2  cm,width=4.3 cm]{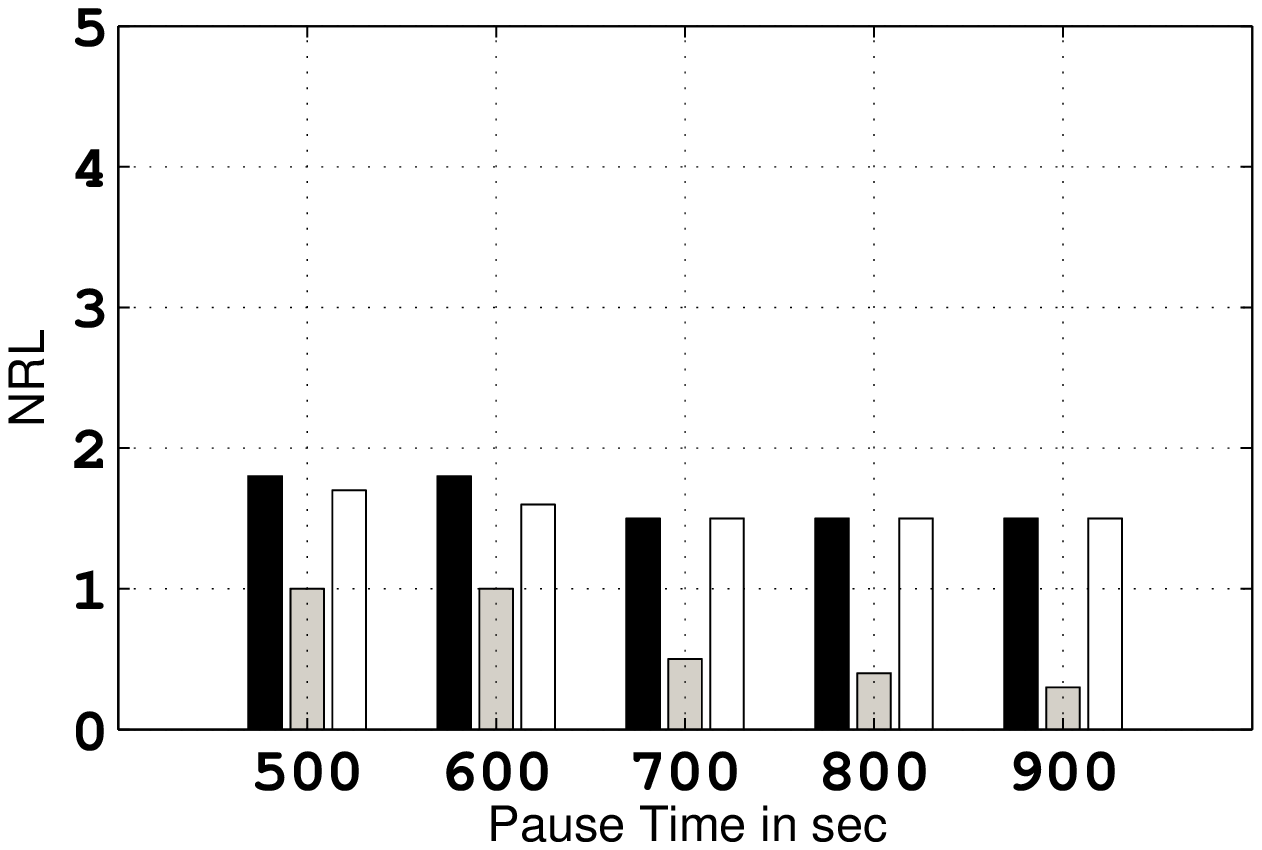}}
  \caption{Simulation results for scenario.2 for varying mobilities}
\end{figure}

\textbf{Scenario.2}
Simulation results for scenario.2 are shown in Fig.3. The area specified is $1000m \times 1000m$ field presenting a square space to allow 50 mobile nodes to move inside. All of the nodes are provided with wireless links of a bandwidth of 2Mbps to transmit on. Simulations are run for 900 seconds each. Each packet starts its journey from a random location and moves towards a random destination with the chosen speed of 15m/s. The 20 source-destination pairs are spread randomly in the network. Once the destination is reached, another random destination is targeted after a specified pause time (from 0s to 900s). A particular scenario for a particular pause time is run for five times and mean of the five obtained values for a particular performance parameter is used to plot the graphs.

\begin{figure}[h]
  \centering
 \subfigure[E2ED for lower scalabilities]{\includegraphics[height=2 cm,width=4.3 cm]{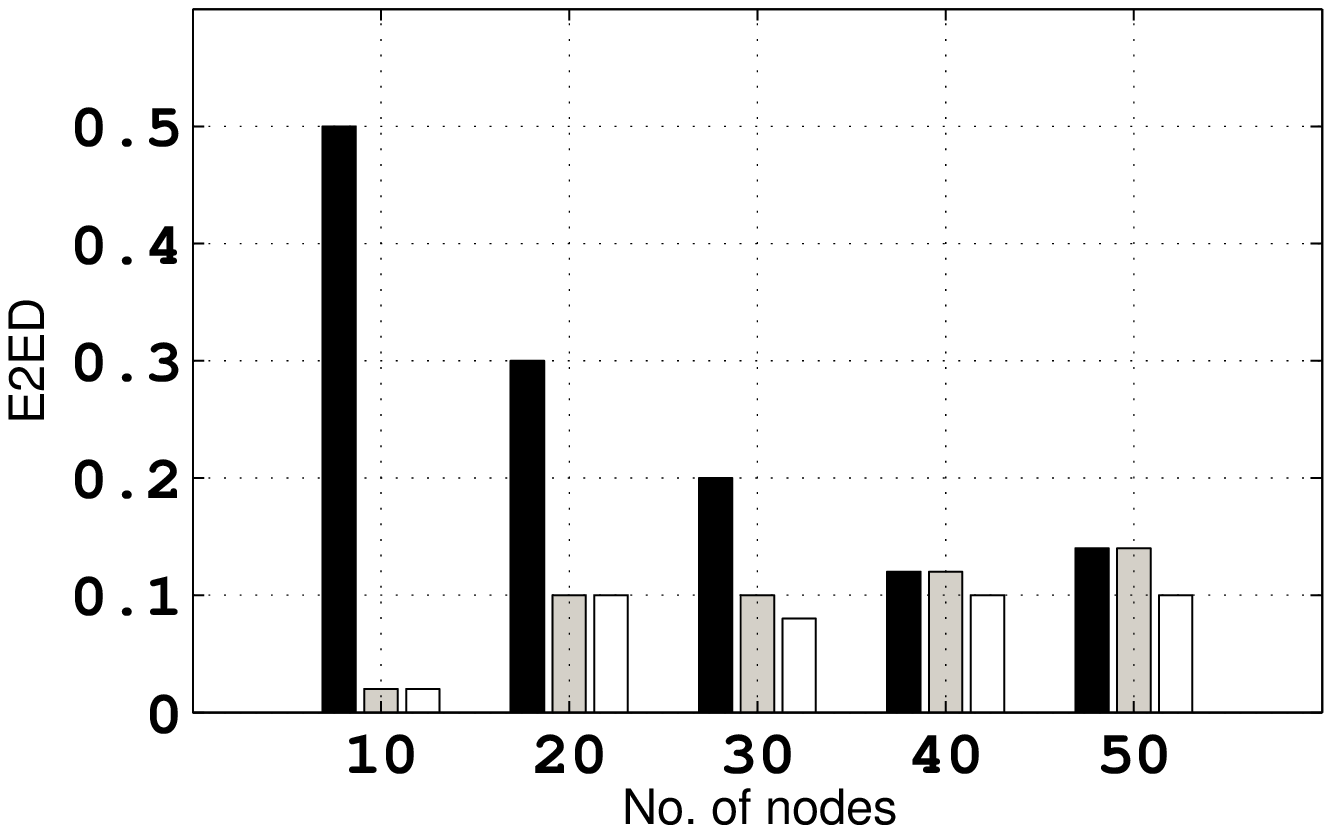}}
 \subfigure[E2ED for higher scalabilities]{\includegraphics[height=2  cm,width=4.3 cm]{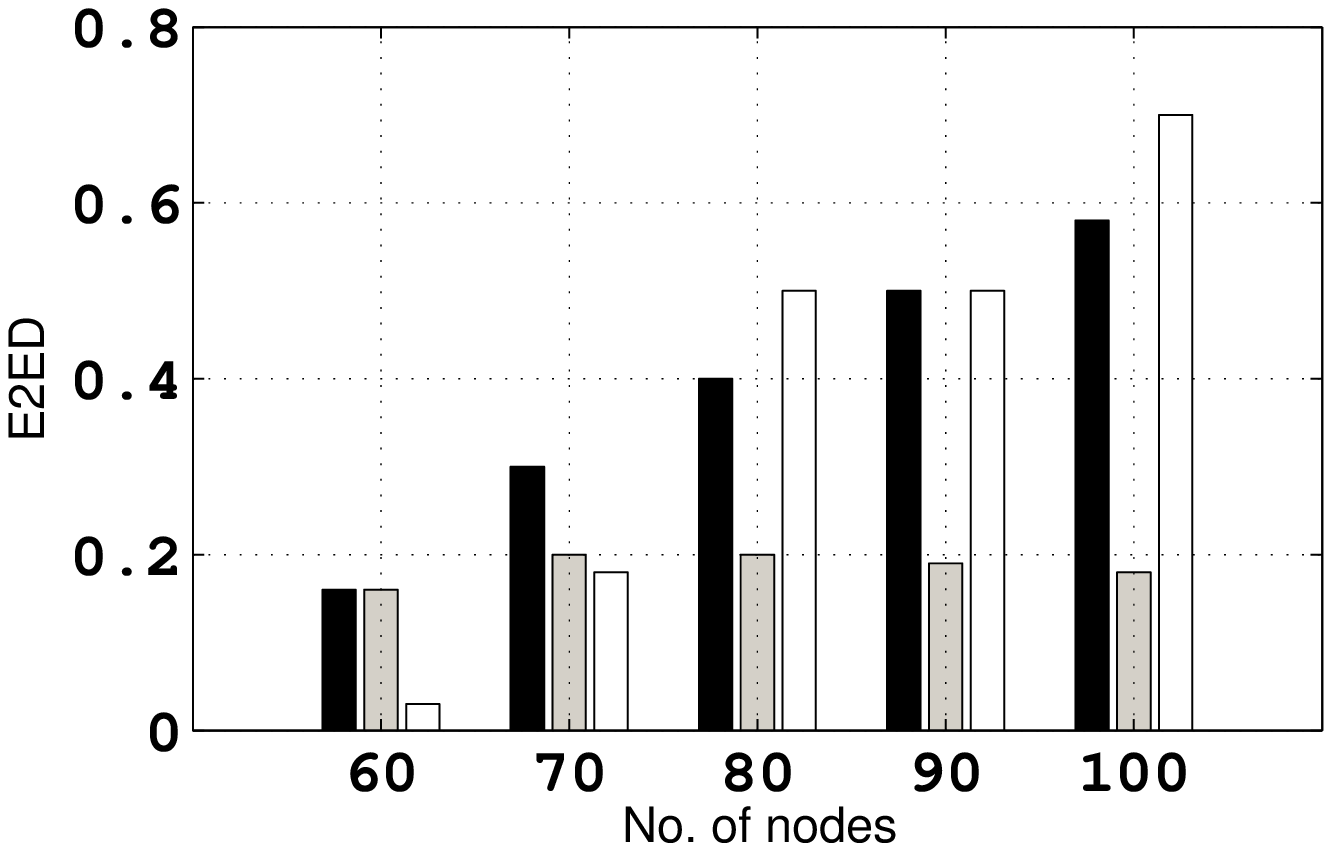}}
 \subfigure[NRL for lower scalabilities]{\includegraphics[height=2 cm,width=4.3 cm]{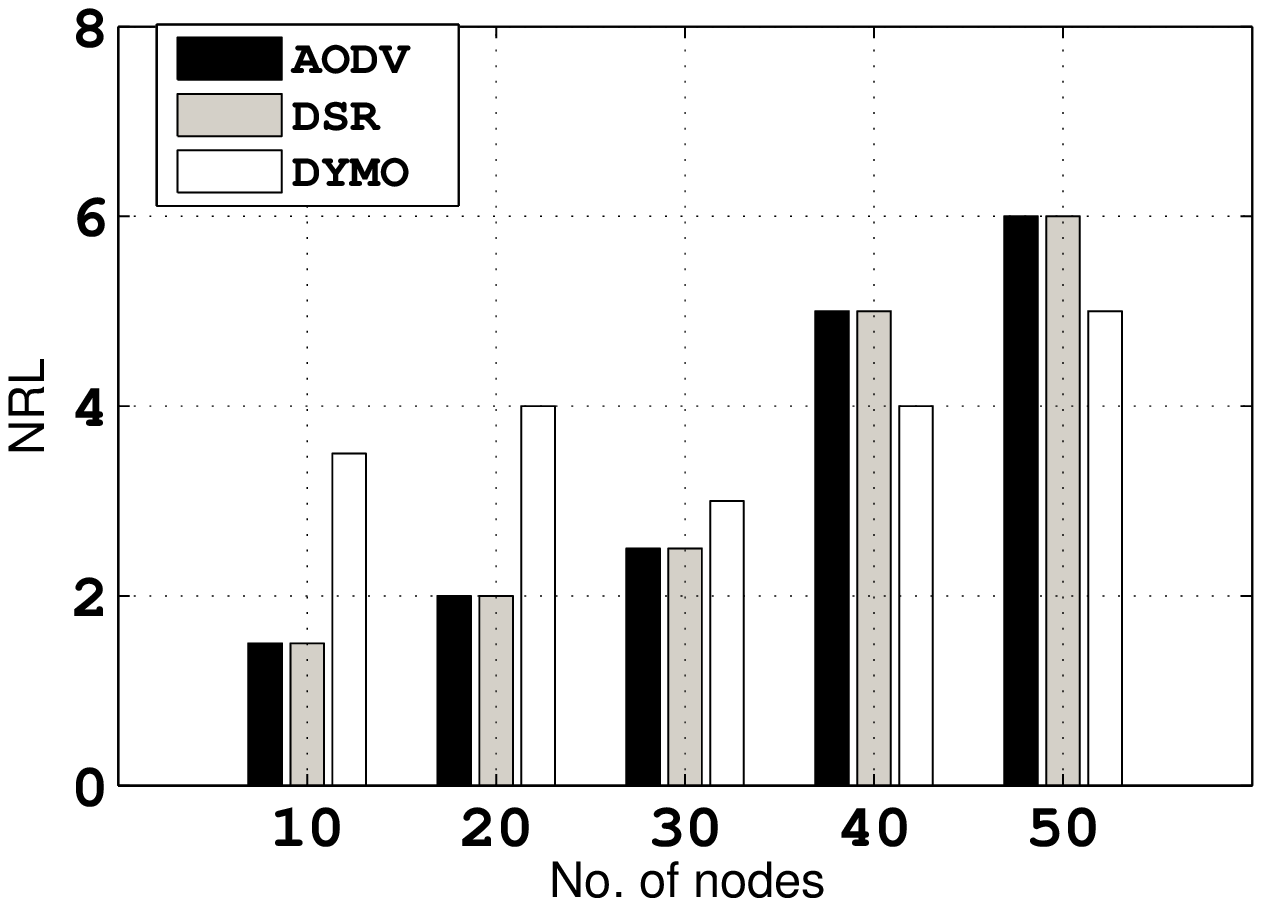}}
 \subfigure[NRL for higher scalabilities]{\includegraphics[height=2  cm,width=4.3 cm]{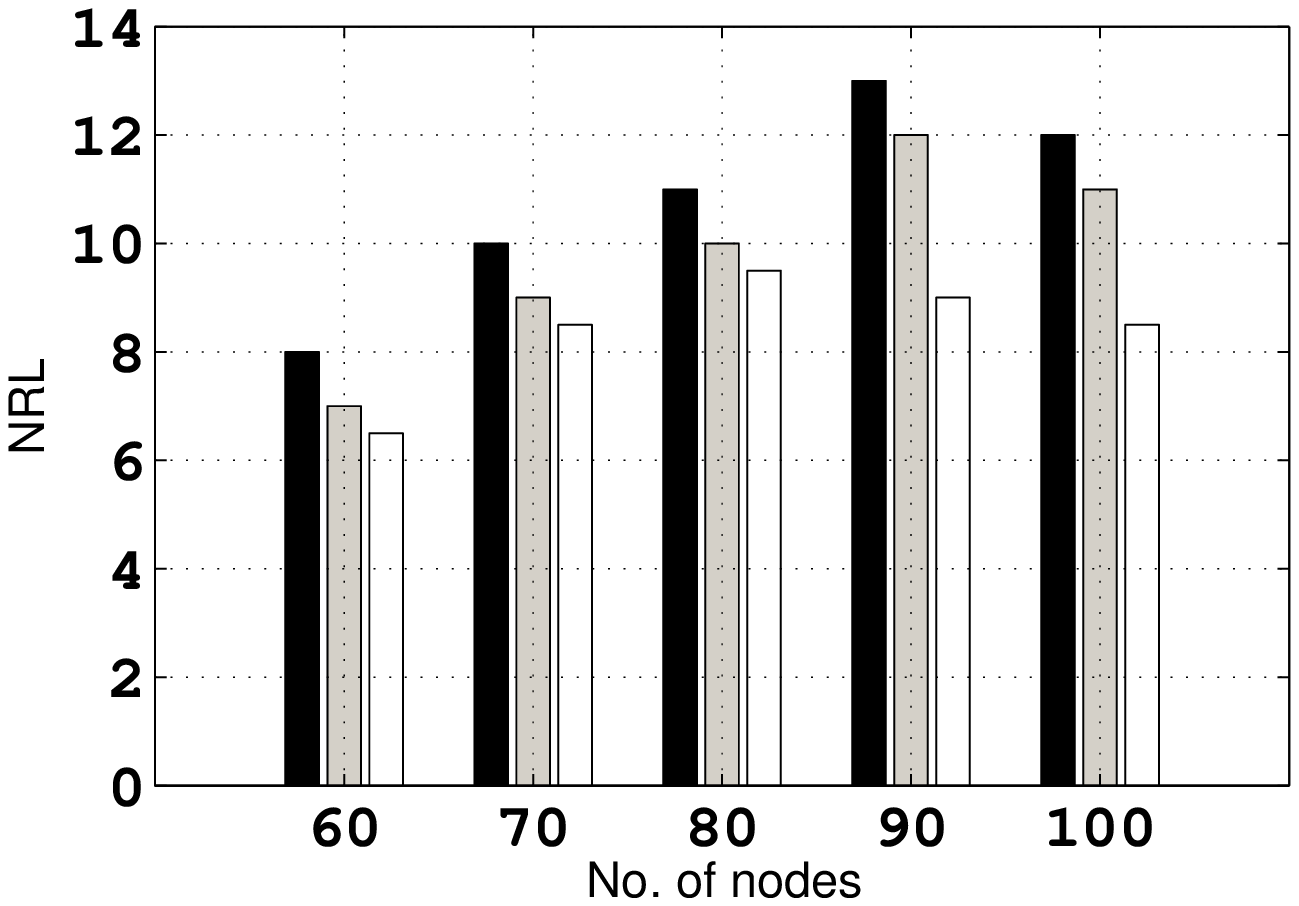}}
  \caption{Simulation results for scenario.3 for varying scalabilities}
\end{figure}

\textbf{Scenario.3}
All of the parameters for Scenario.3 are the same as that of Scenario.2, except a constant pause time of 2s and different scalabilities varying from 10 to 100 nodes. Simulation results are represented in Fig.4.

We have evaluated and compared the protocols by three performance parameters; No. of RREQ Packets, Average End-to-end Delay (E2ED), and Normalized Routing Load (NRL).

\textbf{No. of RREQ Packets in scenario.1.} DYMO generates the highest number of RREQ packets against all RREQ time intervals among all three reactive protocols. This is because of absence of  \textit{grat. RREPs} unlike AODV and DSR. Besides this, any other supplementary routing strategy like $PS$ and $LLR$ is also absent in DYMO. The RREQ retries in DYMO are $3$ which is the highest value of rediscovery retries as compared to AODV and DSR. The promiscuous listening mode of DSR not only helps in $RD$ process by shortening the routes through \textit{grat. RREPs}, but also helps to reduce routing overhead in $RM$ process through $PS$. Therefore, DSR produces less number of RREQ packets both in $5$ hops and $10$ hops in Fig.2.

\textbf{E2ED in scenario.2.} As demonstrated in Fig.3.a,b, AODV among reactive protocols attains the highest delay. Because, high dynamical situations result in frequent link breakage, and $LLR$ operation for $RM$ sometimes results in increased path lengths. DYMO produces the lowest E2ED among reactive protocols because it only uses the ERS for route finding without any supplementary optimization mechanism; as checking $RC$ in DSR and $RT$ in AODV before $RD$ through ERS attains some delay. The reason for consuming more time for end-to-end path establishment in DSR is that for $RD$ it first searches the desired route in $RC$ and then starts $RD$, if the search fails.

\textbf{NRL in scenario.2.} Due to the absence of \textit{grat. RREPs}, (Fig.3.c,d) DYMO produces higher routing overhead. Whereas, DSR, due to the promiscuous listening mode shows the lowest routing load. Although, AODV uses \textit{grat. RREPs} but due to the use of HELLO messages like DYMO and $LLR$, it causes more routing load than DSR. During higher mobilities, (i.e., at low Pause times), the rate of link breakage also increase. In response to this link breakage, all of the on-demand protocols disseminate RERR message to inform the RREQ generator about the faulty links and prevent the use of invalid routes. As a result increased routing overhead occurs.

\textbf{E2ED in scenario.3.} The \textit{grat. RREPs} produce diverse effects in different node density scenarios. DYMO does not use this strategy; the absence of the mechanism keeps the lowest E2ED of DYMO in all scalabilities. First checking the $RC$ instead of simple ERS based $RD$ process augments the delay when population of nodes increases, thus more delay DSR produces which is presented in Fig.4.a,b., as compared to DYMO. AODV experiences the highest E2ED in all scalabilities due to $LLR$ process.

\textbf{NRL in scenario.3.} NRL of DYMO is lowest whereas, the highest one is generated by AODV in all network flows by varying number of nodes, as obvious from Fig.4.c,d. The HELLO messages to check connectivity of active routes, $LLR$ and \textit{grat. RREPs} increase the generation of control packets. Whereas, DSR along with promiscuous listening mode reduces the routing overhead as compared to AODV. Each node participating in $RD$ process (including intermediate nodes) of DSR, learns the routes to other nodes on the route. $PS$ technique is used to get routes from $RC$ of intermediate nodes. This strategy is used to quickly access and to solve broken link issues by providing alternative route.

\vspace{-0.2cm}
\section{Conclusion and Future Work}
In this work, we have modeled the cost paid by 3 reactive protocols; AODV, DSR, and DYMO for the generated routing overhead to keep the topological information updated among wireless nodes. The cost consists of the energy consumed and time spent per packet for route discovery and route maintenance process. Increased routing overhead and increased delay are major issues to be resolved by the routing protocols in wireless environment, so, are considered in this work. Optimizations like $PS$ and $LLR$ of retransmissions result better performance of a protocol. While the reduction of network bandwidth utilization is more useful when data flows are increased. In future, we are interested to re-implement these reactive protocols with new routing link metric; IBETX (Interference and Bandwidth Adjusted Expected Transmission Count) [20] and Inverse ETX [21]. Because, these metrics have achieved reduced overhead by a proactive protocol Destination Sequenced Distance Vector (DSDV).


\end{document}